\documentclass[11pt,a4paper]{article}
\usepackage{amsfonts}
\usepackage{epsfig}
\usepackage{graphicx}
\usepackage{amsmath}
\usepackage{amssymb}
\usepackage{color}
\usepackage[all]{xy}

\begin{document}

\title{Equivalence between Type I Liouville dynamical systems in the plane and the sphere}

\author{M.A. Gonzalez Leon$^1$, J. Mateos Guilarte$^2$ and M. de la Torre Mayado$^2$\\
              \textsl{\small $^1$Departamento de Matem\'atica Aplicada and IUFFyM, University of Salamanca, Spain} \\
              \textsl{\small $^2$Departamento de F\'{\i}sica Fundamental and IUFFyM, University of Salamanca, Spain}}

\date{\small{\today}}

\maketitle

\begin{abstract}
Separable Hamiltonian systems either in sphero-conical coordinates on a $S^2$ sphere or
in elliptic coordinates on a ${\mathbb R}^2$ plane are described in an unified way. A back and forth route connecting these Liouville Type I separable systems is unveiled. It is shown how the gnomonic projection and its inverse map allow us to pass from a Liouville Type I separable system with an spherical configuration space to its Liouville Type I partner where the configuration space is a plane and back. Several selected spherical separable systems and their planar cousins are discussed in a classical context.
\end{abstract}

\section{Introduction}
\label{intro}

Hamiltonian systems in ${\mathbb R}^2$ that admit separation of variables were completely determined by Liouville \cite{Liouville1859} and Morera \cite{Morera}, and can be classified, see \cite{Perelomov}, in four different types according with the system of coordinates where the separability is manifested: elliptic, polar, parabolic and Cartesian respectively. Thus Type I Liouville systems in ${\mathbb R}^2$ are defined by natural Hamiltonians: $H=K+{\cal U}$, $K=\frac{m}{2} \left( (\frac{dx_1}{dt})^2+(\frac{dx_2}{dt})^2\right)$, such that in elliptic coordinates adopt the Liouville form \cite{Perelomov}.

In this work we shall establish an isomorphism between this kind of mechanical systems and Liouville systems in $S^2$ that are separable in sphero-conical coordinates, that, correspondingly, we shall call Type I Liouville systems in $S^2$.

This isomorphism will be constructed by mapping the configuration space $S^2$ by means of two gnomonic projections from the two $S^2$-hemispheres into two ${\mathbb R}^2$ planes, together with a redefinition of the physical time and the application of a linear transformation in the projecting planes. This procedure is a generalization of the method used in \cite{Gonzalez}, where the orbits of the two fixed center problem on $S^2$ \cite{Killing1885,Kozlov1992} were determined by inverting these transformations. The inspiration was taken from the work of Borisov and Mamaev \cite{Borisov2007}, based itself on the ideas of Albouy \cite{Al1,Al2}, the main novelty of \cite{Gonzalez} was the simultaneous consideration of two gnomonic projections in order to study the complete set of orbits, identifying each trajectory crossing the equator of $S^2$ with the conjunction of two planar unbounded orbits, one of the two attractive center problem and another corresponding to the system of the two associated repulsive centers.

The idea of projecting dynamical systems in constant positive curvature surfaces to planar ones goes back to Appell \cite{Appell1890,Appell1891} and has been developed in modern times by Albouy \cite{Al1,Al2,Al3,Albouy2013,Al4}, see also \cite{Borisov2016} for a detailed historical review and references on problems defined in spaces of constant curvature.

The structure of this paper is as follows: In Section 2 the gnomonic projections will be constructed. Section 3 is devoted to describe the properties of Liouville type I systems, both in $S^2$ and ${\mathbb R}^2$. In Section 4 the isomorphism is established. Section 5 contains several selected examples and finally some comments and future perspectives are showed in the final section.

\section{Gnomonic projections from $S^2$ to ${\mathbb R}^2$}
\label{sec:1}

Let us consider the $S^2$ sphere embedded in ${\mathbb R}^3$, i.e. $(X,Y,Z)\in {\mathbb R}^3$, such that $X^2+Y^2+Z^2=R^2$. Standard spherical coordinates in $S^2$:
\[
X=R \sin \theta \cos \varphi\ ,\quad Y=R \sin \theta \sin \varphi \  ,\quad Z=R \cos \theta
\]
$\theta\in [0,\pi]$, $\varphi\in [0,2\pi)$, allow us to write the metric tensor in $TS^2$ (i.e. the restriction of Euclidean metric in $T{\mathbb R}^3$ to the sphere) in standard form:
\[
ds^2=R^2 \left( d\theta^2+\sin^2\theta \, d\varphi^2\right)\label{metrica1}
\]
The gnomonic projections from the North/South hemispheres: $S^2_+=\{ (X,Y,Z)\in S^2/ Z>0\}$, $S^2_-=\{ (X,Y,Z)\in S^2/ Z<0\}$, to the ${\mathbb R}^2$ plane, with respect to the points $(0,0,\pm R)$, are defined by the change of variables
\[
\Pi_\pm: S_\pm^2 \longrightarrow {\mathbb R}^2\quad \Rightarrow \quad \left\{ \begin{array}{ll}  x=\frac{R}{Z} X = R \tan \theta \cos \varphi \\ & \\ y=\frac{R}{Z} Y =R \tan \theta \sin \varphi \end{array}\right.\  ,\quad  \varphi\in[0,2\pi)
\]
where $\theta \in \left[ 0, \frac{\pi}{2} \right)$ in the case of $\Pi_+$ and $\theta \in   \left( \frac{\pi}{2},\pi\right] $ for $\Pi_-$. The inverse maps $\Pi_{\pm}^{-1}: {\mathbb R}^2 \longrightarrow S_{\pm}^2$, read:
\[
X=\frac{Rx}{\sqrt{R^2+x^2+y^2}}\  ,\  Y=\frac{Ry}{\sqrt{R^2+x^2+y^2}}\  ,\  Z=\frac{\pm R^2}{\sqrt{R^2+x^2+y^2}}
\]

The projections $\Pi_{\pm}$ define two copies of the Riemannian manifold $({\mathbb R}^2,g)$ where the metric tensor $g$ in each copy is given by:
\begin{equation}
ds^2=\frac{R^2}{(R^2+x^2+y^2)^2} \left( (R^2+y^2) dx^2 -2xy\, dx \, dy +(R^2+x^2) dy^2\right)\label{metrica2}
\end{equation}
with associated Christoffel symbols: $ \Gamma_{22}^1= \Gamma_{11}^2=0$,
\[
\Gamma_{11}^1=2\Gamma_{12}^2=2\Gamma_{21}^2= \frac{-2x}{R^2+x^2+y^2}\, ,\quad  \Gamma_{22}^2=2\Gamma_{12}^1=2\Gamma_{21}^1= \frac{-2y}{R^2+x^2+y^2}\  \  .
\]

\begin{figure}
\begin{center}  \includegraphics[height=7cm]{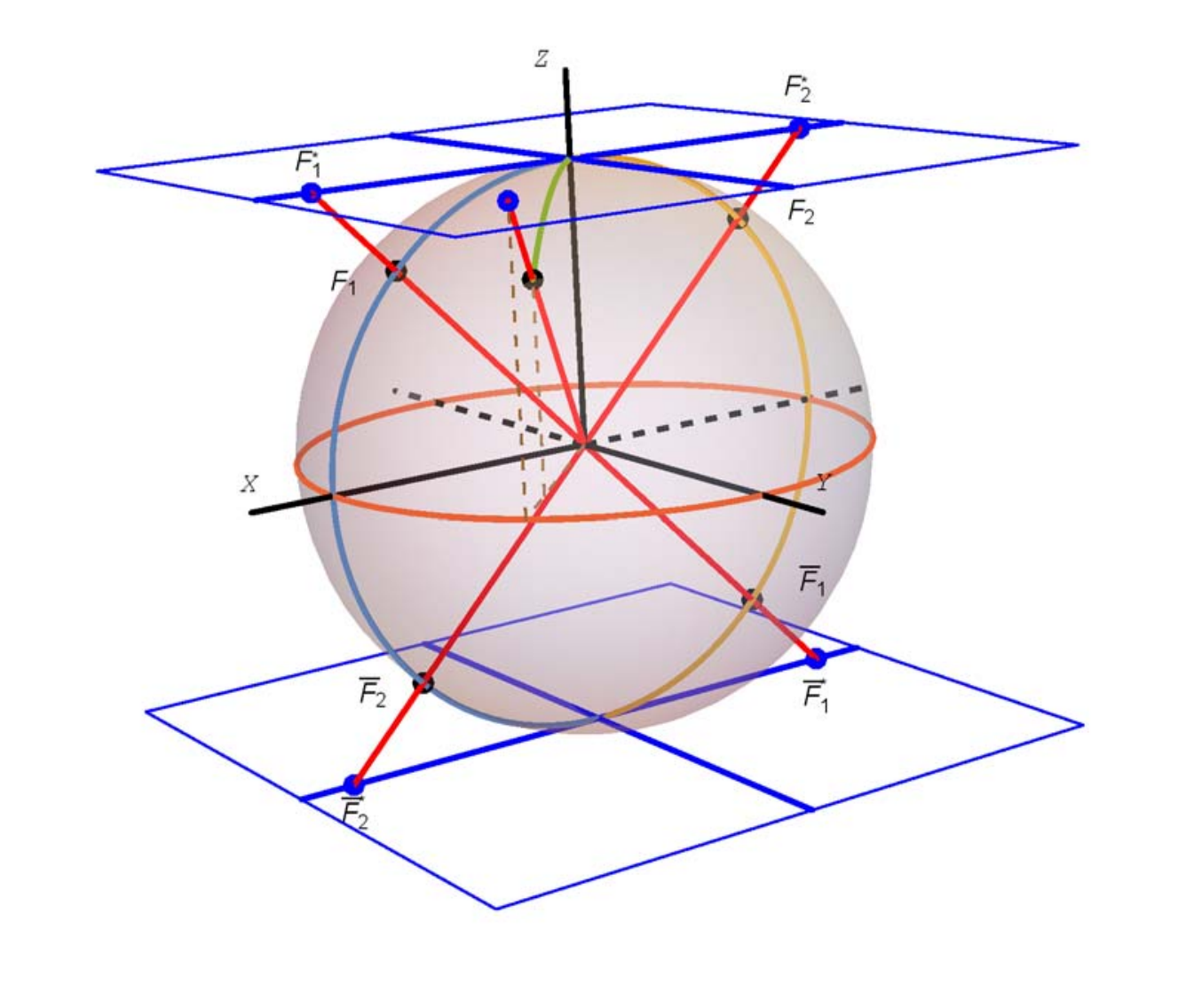}
\caption{Gnomonic projections $\Pi_+$ and $\Pi_-$.}
\end{center}
\label{fig:1}
\end{figure}

Gnomonic projections map geodesics in $S^2$ into straight lines in ${\mathbb R}^2$. In fact the geodesic equations for the metric (\ref{metrica2}):
\[
\nabla_{\dot{\bf x}} {\dot{\bf x}} =0 \Rightarrow \left\{ \begin{array}{lcc} \ddot{x}+\Gamma_{11}^1 \dot{x}^2+2\Gamma_{12}^1 \dot{x}\dot{y} +\Gamma_{22}^1 \dot{y}^2 &=& 0\\ \ddot{y}+\Gamma_{11}^2 \dot{x}^2+2\Gamma_{12}^2 \dot{x}\dot{y} +\Gamma_{22}^2 \dot{y}^2 &=& 0 \end{array}\right.
\]
where ${\bf x}\equiv (x(t),y(t))\in {\mathbb R}^2$ and dots represent derivative with respect to $t$, can be converted, by changing from physical to local (or \textit{projected}) time, into trivial standard form:
\[
d\tau = \frac{R^2+x^2+y^2}{R^2} dt \quad  \Rightarrow \quad   x^{\prime\prime}=0\  ,\quad y^{\prime\prime}=0
\]
where primes denote derivation with respect to $\tau$.

Given a mechanical problem in $S^2$ defined by a potential function ${\cal U}$, the projection of Newton equations in $S^2_+$ or $S^2_-$ to $({\mathbb R}^2,g)$ can be written as:
\begin{equation}
\nabla_{\dot{\bf x}} {\dot{\bf x}} =-{\rm grad} \, {\cal U}({\bf x})\label{equation1}
\end{equation}
where ${\bf x}\equiv (x,y)$, and covariant derivatives and the gradient are associated to the $g$ metric (\ref{metrica2}). Changing to projected time, equations (\ref{equation1}) will be written as:
\begin{equation}
{\bf x}'' = -{\rm grad} \, {\cal U}({\bf x})\Rightarrow \left\{ \begin{array}{lcl} x^{\prime\prime}&=&-g^{11}\frac{\partial{\cal U}}{\partial x}-g^{12}\frac{\partial{\cal U}}{\partial y} \\ y^{\prime\prime}&=& -g^{21}\frac{\partial{\cal U}}{\partial x}-g^{22}\frac{\partial{\cal U}}{\partial y}\end{array} \right. \label{newton}
\end{equation}
where $g^{ij}$ denote the components of $g^{-1}$, the inverse of the metric $g$.

We now pose the following question: Is it possible to understand equations (\ref{newton}) as Newton equations for a mechanical system in the Euclidean ${\mathbb R}^2$ plane with time $\tau$?, in other words: Would it exists a function ${\cal V}(x_1,x_2)$ such that equations
\begin{equation}
x_1^{\prime\prime}=-\frac{\partial{\cal V}}{\partial x_1} \quad , \quad x_2^{\prime\prime}=-\frac{\partial{\cal V}}{\partial x_2} \label{newton1}
\end{equation}
are equivalent to (\ref{newton})?

The answer was given by Albouy \cite{Al1} and developed explicitly by Borisov and Mamaev \cite{Borisov2007} for the case of the Killing problem restricted to the North Hemisphere, i.e. the problem of two Kepler centers in $S^2_+$. The equivalence (trajectory isomorphism) was achieved in this concrete case via the linear transformation $x_1=x$, $x_2=\frac{1}{\sigma}y$, for an adequate value of $\sigma$ parameter, in equations (\ref{newton}).  Moreover, in \cite{Borisov2007} this isomorphism was extended to other mechanical systems and in general to systems admitting separation of variables in sphero-conical coordinates in $S^2_+$. In \cite{Gonzalez} the equivalence for the Killing problem was applied to the complete sphere considering the two projections $\Pi_+$ and $\Pi_-$ simultaneously. A delicate point is the gluing of the inverse projections at the equator of the sphere. Orbits crossing the equator have to be described by the differentiable gluing of two pieces coming from unbounded orbits in each of the two planes respectively.

In this work, following \cite{Borisov2007}, we shall show that these results are valid for the class of Type I Liouville systems in the whole $S^2$, i.e. separable system in sphero-conical coordinates in $S^2$, that are transformed by gnomonic projections and the linear transformation, into Liouville systems of type I in ${\mathbb R}^2$ (separable in elliptic coordinates) with respect to the ``non-physical" time $\tau$.

\section{Liouville type I systems in $S^2$ and ${\mathbb R}^2$}

We shall refer to Hamilton-Jacobi separable spherical systems in sphero-conical coordinate as Liouville dynamical systems of Type I in $S^2$, in analogy with the planar case for elliptic coordinates, see \cite{Perelomov},

Sphero-conical coordinates $U\in(\bar\sigma,1)$, $V\in (-\bar\sigma,\bar\sigma)$ describe points in an $S^2$-sphere by means of the geodesic distances $R\theta_1$ and $R\theta_2$ from the particle position to two fixed points that we choose without loosing generality as: $F_1=(R\bar{\sigma},0,R\sigma)$, $F_2=(-R\bar{\sigma},0,R\sigma)$, $\sigma=\cos \theta_f$, $\bar{\sigma}=\sin \theta_f$, see Figure 2, in the form:
\[
\theta_1={\rm arccos}({\sigma \cos\theta+\bar{\sigma}\sin\theta\cos\varphi)}\  ,\quad \theta_2={\rm arccos}({\sigma \cos\theta-\bar{\sigma}\sin\theta\cos\varphi})
\]
Sphero-conical coordinates are thus defined by replicating on the sphere the \lq\lq gardener\rq\rq \   construction which allowed Euler to define elliptic coordinates in ${\mathbb R}^2$:
\[
U=\sin\frac{\theta_1+\theta_2}{2}\  ,\quad V=\sin\frac{\theta_2-\theta_1}{2}
\]
and the change of coordinates is the following:
\[
X=\frac{R}{\bar{\sigma}} UV\  ,\quad Y^2=\frac{R^2}{\sigma^2\bar{\sigma}^2} (U^2-\bar{\sigma}^2)(\bar{\sigma}^2-V^2)\  ,\quad Z^2=\frac{R^2}{\sigma^2} (1-U^2)(1-V^2) \  .
\]

\begin{figure}
\begin{center}
\includegraphics[height=5cm]{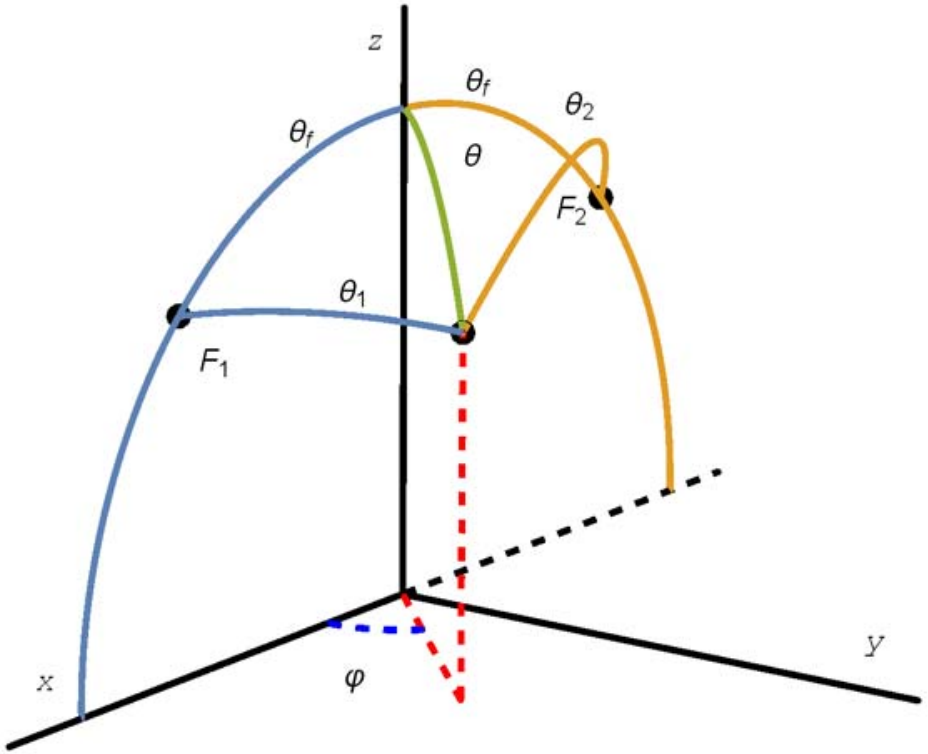}
\caption{Position of the particle in the sphere relative to two fixed points or foci.}
\end{center}
\label{fig:1}       
\end{figure}

The kinetic energy of a particle moving on one $S^2$ sphere as configuration space expressed in sphero-conical coordinates reads:
\[
K=\frac{mR^2}{2}\, \left[ \frac{U^2-V^2}{(1-U^2)(U^2-\bar{\sigma}^2)} \left( \frac{dU}{dt}\right)^2 +\frac{U^2-V^2}{(1-V^2)(\bar{\sigma}^2-V^2)} \left( \frac{dV}{dt}\right)^2 \right] \  ,
\]
where $t$ is the physical time and we stress that $K$ is singular in the Equator, i.e., in the circle: $Z=0 \, \equiv \, U=1$.
Changing from physical to local time, $d\varsigma = \frac{dt}{U^2-V^2}$, the Kinetic energy is rewritten as:
\[
K=\frac{m R^2}{2(U^2-V^2)} \left[ \frac{1}{(1-U^2)(U^2-\bar{\sigma}^2)} \left( \frac{dU}{d\varsigma}\right)^2 +\frac{1}{(1-V^2)(\bar{\sigma}^2-V^2)} \left( \frac{dV}{d\varsigma}\right)^2 \right]\, .
\]
We define a natural dynamical system as Liouville of Type I in $S^2$ if the potential energy is a function of the form:
\begin{equation}
{\cal U}(U,V) = \frac{1}{U^2-V^2} \left( F(U)+G(V) \right) \quad .\label{typeIS2}
\end{equation}
This kind of potentials where the functions $F(U)$ and $G(V)$ are regular enough give rise to motion equations which are separable in the $U$ and $V$ evolutions.

Systems of this type are automatically completely integrable. The first integral of motion, the mechanical energy $E=K+{\cal U}$, leads to the separated expressions:
\[
 - \frac{m R^2\left( \frac{dV}{d\varsigma}\right)^2}{2(1-V^2)(\bar{\sigma}^2-V^2)}\, -G(V) -E V^2  \, =\,  \frac{m R^2 \left( \frac{dU}{d\varsigma}\right)^2}{2(1-U^2)(U^2-\bar{\sigma}^2)}\, +F(U) - E U^2
 \]
which necessarily must be equal to a constant $-\Omega$, a second invariant in involution with the energy. Rearranging these expressions we finally reduce the equations of motion to the uncoupled first-order ODE's system:
\begin{eqnarray}
\left( \frac{dU}{d\varsigma}\right)^2 &=& \frac{2}{mR^2}\   (1-U^2)(U^2-\bar{\sigma}^2)\,  (-\Omega+E \, U^2 - F(U)) \label{ode} \\ \left( \frac{dV}{d\varsigma}\right)^2 &=& \frac{2}{mR^2}\   (1-V^2)(\bar{\sigma}^2-V^2)\,  (\Omega-E \, V^2 - G(V)) \label{vode} \quad .
\end{eqnarray}
that is immediately integrated via the quadratures:
\begin{eqnarray}
\varsigma -\varsigma_0 &=& \pm R\sqrt{\frac{m}{2}}\int_{\bar\sigma}^U \frac{d \tilde{U}}{\sqrt{(1-\tilde{U}^2)(\tilde{U}^2-\bar{\sigma}^2)\,  (-\Omega+E \, \tilde{U}^2 - F(\tilde{U}))}} \label{quadrature1} \\ \nonumber
\\
\varsigma -\varsigma_0 &=& \pm R\sqrt{\frac{m}{2}}\int_{-\bar\sigma}^V \frac{d \tilde{V}}{\sqrt{ (1-\tilde{V}^2)(\bar{\sigma}^2-\tilde{V}^2)\,  (\Omega-E \, \tilde{V}^2 - G(\tilde{V}))}} \quad . \label{quadrature2}
\end{eqnarray}
and the orbits are found by inversion, if possible, of these integrals. The physical time can be recovered by integration the expression:
\[
t=\int_{\varsigma_0}^{\varsigma} (U(\bar{\varsigma})^2-V(\bar{\varsigma})^2) d\bar{\varsigma}
\]

Liouville Type I systems in ${\mathbb R}^2$ are separable in elliptic coordinates \cite{Perelomov}. Recall that Euler elliptic coordinates in ${\mathbb R}^2$ relative to the foci: $f_1=(a,0)$, $f_2=(-a,0)$ are defined as half the sum and half the difference of the distances from the particle position to the foci:
\begin{equation}
 u=\frac{r_1+r_2}{2a } \, , \   v=\frac{r_2-r_1}{2 a}
 \, ; \   r_1=\sqrt{(x_1-a)^2+x_2^2} \, , \  r_2=\sqrt{(x_1+a)^2+x_2^2} \  . \label{euler}
\end{equation}
The new coordinates vary in the intervals: $-1<v<1$, $1<u<\infty$. In terms of these coordinates the particle position is defined to be
\begin{equation}
 x_1=a  uv \ ,\quad x_2^2= a^2\,  (u^2-1)(1-v^2) \quad ,\label{elliptic}
\end{equation}
implying a two-to-one map from ${\mathbb R}^2$ to the infinite ``rectangle": $(-1,1)\times (1, +\infty)$. The Kinetic energy with respect to the local time $d\zeta = \frac{dt}{u^2-v^2}$ in this coordinate system reads
\[
K=\frac{m a^2}{2(u^2-v^2)}\, \left( \frac{1}{u^2-1} \left( \frac{du}{d\zeta}\right)^2 +\frac{1}{1-v^2} \left( \frac{dv}{d\zeta}\right)^2 \right)
\]
and the potential provides a Liouville Type I system in ${\mathbb R}^2$ if is of the form
\begin{equation}
{\cal V}(u,v) = \frac{1}{u^2-v^2} \left( f(u)+g(v) \right) \  ,\label{sepellip}
\end{equation}
for arbitrary but sufficiently regular functions $f(u)$ and $g(v)$. An standard separability process leads to the uncoupled first-order ODE's:
\begin{eqnarray}
\left( \frac{du}{d\zeta}\right)^2 &=& \frac{2}{ma^2}\   (u^2-1)\,  (-\lambda+h \, u^2 -f(u)) \label{ueq}\\
\left( \frac{dv}{d\zeta}\right)^2 &=& \frac{2}{ma^2}\   (1-v^2)\,  (\lambda-h \, v^2 - g(v))\label{veq}
\end{eqnarray}
depending on the energy $h$ and the second constant of motion $\lambda$.

\section{Trajectory Isomorphism between Liouville Type I systems in $S^2$ and ${\mathbb R}^2$}

The gnomonic projection $\Pi_+$ from $S^2_+$ to ${\mathbb R}^2$ allows us to write the cartesian coordinates $(x,y)$ in terms of the sphero-conical ones:
\[
x=\frac{R\sigma}{\bar{\sigma}} \, \frac{UV}{\sqrt{1-U^2}\sqrt{1-V^2}}\  ,\quad y^2=\frac{R^2}{\bar{\sigma}^2} \, \frac{(U^2-\bar{\sigma}^2)(\bar{\sigma}^2-V^2)}{(1-U^2)(1-V^2)}
\]
The re-scaling $x_1=x$, $x_2=\frac{y}{\sigma}$ permits us to re-write these expressions in terms of Euler elliptic coordinates in the form (\ref{elliptic}) for $a=\frac{R\bar{\sigma}}{\sigma}$. Note that with this choice we have that $\Pi_+(F_1)=(a,0)=f_1$ and $\Pi_+(F_2)=(-a,0)=f_2$. Thus the ${\mathbb R}^2$ plane with coordinates $(x_1,x_2)$ is equivalently described in terms of the sphero-conical coordinates or in the elliptic form (\ref{elliptic}) via the identifications:
\begin{equation}
u=\frac{\sigma U}{\bar{\sigma} \sqrt{1-U^2}}\  , \quad v=\frac{\sigma V}{\bar{\sigma} \sqrt{1-V^2}}\label{identification1}
\end{equation}

Let us consider a Liouville Type I system in $S^2$ with potential (\ref{typeIS2}) and the corresponding separated first-order equations (\ref{ode},\ref{vode}) with respect to the local time $\varsigma$ in $S^2$. The chain of changes from this local time $\varsigma$ to the elliptic local time $\zeta$, via going back to the physical time $t$, changing to the projected time $\tau$ and finally defining $\zeta$ in terms of $\tau$: $d\zeta = \frac{d\tau}{u^2-v^2}$, can be simply summarized in the form:
\[
d\zeta= \bar{\sigma}\, d\varsigma
\]
Using this expression and the identification (\ref{identification1}) one easily realizes that equations (\ref{ode},\ref{vode}) are equivalent to equations (\ref{ueq},\ref{veq}), and reciprocally, if the respective constants of motion are related through the equation:
\[
h=\frac{E-\Omega}{\sigma^2} \quad , \quad \lambda =\frac{\Omega}{\bar{\sigma}}  \quad .
\]
and the the potential energy in ${\mathbb R}^2$ is obtained from the potential energy in $S^2$, and viceversa,
via the identities:
\begin{equation}
f(u)=\frac{\bar{\sigma}^2 u^2+\sigma^2}{\sigma^2 \bar{\sigma}^2}\, F(U(u))\   , \quad
g(v)=\frac{\bar{\sigma}^2 v^2+\sigma^2}{\sigma^2 \bar{\sigma}^2}\, G(V(v)) \label{fyg}
\end{equation}
It is thus established the prescription to pass back and forth between a Liouville Type I separable systems in $S^2_+$ with a given physical time $t$ and a Liouville Type I system in ${\mathbb R}^2$ with respect to a ``non-physical" time $\tau$.

An analogous procedure relative to the projection $\Pi_-$ can be developed for $S^2_-$, and thus Newton equations on $S^2_{\pm}$ are equivalent to Newton equations (\ref{newton1}) in the Euclidean planes. The orbits of a system with $S^2$ as configuration space require the determination of the orbits of two planar systems to be completely described in this projected picture.

In order to clarify the relationship between local times needed for separability in $S^2$ and $R^2$ we include a Table showing all the changes of time schedules:
\begin{center}
\entrymodifiers={++[F]}
\xymatrix@R=1.0cm{
{\begin{array}{c}  \txt{\scriptsize $S^2_\pm$, Newton Equations for ${\cal U}$}     \\  \txt{\scriptsize Physical time $t$} \end{array}}  \ar[d]^{\txt {\scriptsize Sphero-Conical coor.}} \ar[r]^{\txt{\scriptsize Gnomonic Proj.}} &
{\begin{array}{c}  \txt{\scriptsize $({\mathbb R}^2,g)$, Newton Eqs. for ${\cal U}$}     \\  \txt{\scriptsize Physical time $t$} \end{array}}    \ar[d]^{\txt{\scriptsize ``Projected" time $\tau$}}\\ {\begin{array}{c}  \txt{\scriptsize Separable problem in $S^2_\pm$}  \\  \txt{\scriptsize Physical time $t$} \end{array}} \ar[d]^{\txt{\scriptsize Local time $\varsigma$}}  & {  \txt{\scriptsize ${\mathbb R}^2$, System of second order ODE. Projected time $\tau$} }  \ar[d]^{\txt{\scriptsize Linear transf.}} \\    { {\begin{array}{c}   \txt{\scriptsize Separated First Order Eqs. in $S^2_\pm$}     \\  \txt{\scriptsize Local time $\varsigma$} \end{array}}} \ar@2{<->}[ddr]  &  {    \txt{\scriptsize  $({\mathbb R}^2,\delta_{ij})$, Newton Eqs. for ${\cal V}$. Time $\tau$} }\ar[d]^{\txt {\scriptsize Elliptic coor.}}\\ * { } & {  \txt{\scriptsize Separable problem in ${\mathbb R}^2$. Time $\tau$} } \ar[d]^{\txt{\scriptsize Local time $\zeta$}} \\  {\text{\scriptsize $d\zeta = \bar{\sigma} \, d\varsigma$ }  } & {\begin{array}{c}   \txt{\scriptsize Separated First Order Eqs. in ${\mathbb R}^2$}     \\  \txt{\scriptsize Projected-Local time $\zeta$} \end{array}}
}
\end{center}

\section{Gallery of selected examples}

\subsection{The Neumann system}
The Neumann system \cite{Neumann} consists of a particle constrained to move in a $S^2$ sphere of radius $R$ subjected to maximally anisotropic linear attraction towards the center of the sphere. The potential energy is:
\begin{equation}
{\cal U}(X,Y,Z)= a X^2+b Y^2+c Z^2 \  , \quad a>b>c>0 \quad , \label{Neumann}
\end{equation}
where the couplings $a,b,c$ may be redefined as $\frac{m \omega^2}{2}=a-c$, $0<\sigma^2=\frac{b-c}{a-c}<1$, to easily show that the Neumann
problem is a Liouville Type I system in $S^2$ since the potential energy in sphero-conical coordinates is of the standard form (\ref{typeIS2}) with:
\[
F(U)=\frac{m\omega^2}{2}R^2\left(U^2-\bar{\sigma}^2\right)U^2 \   , \quad G(V)=\frac{m\omega^2}{2}R^2\left(\bar{\sigma}^2-V^2\right) V^2  \quad .
\]
The sigma parameter fixing the position of the foci measures in this case the asymmetry between the intensity of the elastic forces in the $X$ and $Y$ directions. Consequently, the orbits of a particle in the Neumann problem are determined by evaluating the quadratures (\ref{quadrature1},\ref{quadrature2}) with this choice of $F(U)$ and $G(V)$. Both integrals can be written in the compact form:
\begin{eqnarray}
&& \varsigma-\varsigma_0
=\pm \sqrt{m}R \int_{{\cal X}_0}^{\cal X} \, \frac{d \tilde{{\cal X}}}{\sqrt{P_5(\tilde{{\cal X}})}} \label{hyperell} \\ && \hspace{-1cm}
P_5({\cal X})=-{\cal X}(1-{\cal X})({\cal X}-\bar{\sigma}^2)\left(m\omega^2 R^2{\cal X}^2-(2E -m \omega^2R^2\bar{\sigma}^2){\cal X}+2\Omega\right) \nonumber
\end{eqnarray}
where a new integration variable: ${\cal X}$ has been introduced: $U={\cal X}^2$ for quadrature (\ref{quadrature1}) and $V={\cal X}^2$ for (\ref{quadrature2}). (\ref{hyperell}) is an hyperelliptic integral of genus 2, and obviously to obtain explicit expressions for these orbits requires the use of rank 2 Theta functions, see \cite{Dubrovine}, \cite{Enolski}.

Having into account the symmetry of the Neumann potential (\ref{Neumann}), the corresponding planar potentials ${\cal V}(x_1,x_2)$ will have identical expressions in both $\Pi_+(S^2_+)$ and $\Pi_-(S^2_-)$ planes. Applying (\ref{fyg}) we obtain potential (\ref{sepellip}) with:
\[
f(u)=\frac{m\omega^2R^2\sigma^2}{2}\, \frac{u^2(u^2-1)}{\bar{\sigma}^2 u^2+\sigma^2}\  ,\quad g(v)=\frac{m\omega^2R^2\sigma^2}{2} \  \frac{v^2(1- v^2)}{\bar{\sigma}^2 v^2+\sigma^2}
\]
that in cartesian coordinates corresponds to the  potential function:
\begin{equation}
{\cal V}(x_1,x_2)= \frac{m\omega^2 R^2}{2} \left( \frac{x_1^2+\sigma^2 x_2^2}{R^2+x_1^2+\sigma^2 x_2^2} \right) \label{Neumannplano}
\end{equation}
Thus orbits for (\ref{Neumann}) lying in $S^2_+$ or $S^2_-$ are in one to one correspondence with bounded orbits of the planar system (\ref{Neumannplano}) whereas orbits that crosses the equator have to be recovered from the projected pictures as the gluing of unbounded orbits of the two planar copies.

\begin{figure*}
\begin{center}
\includegraphics[height=4cm]{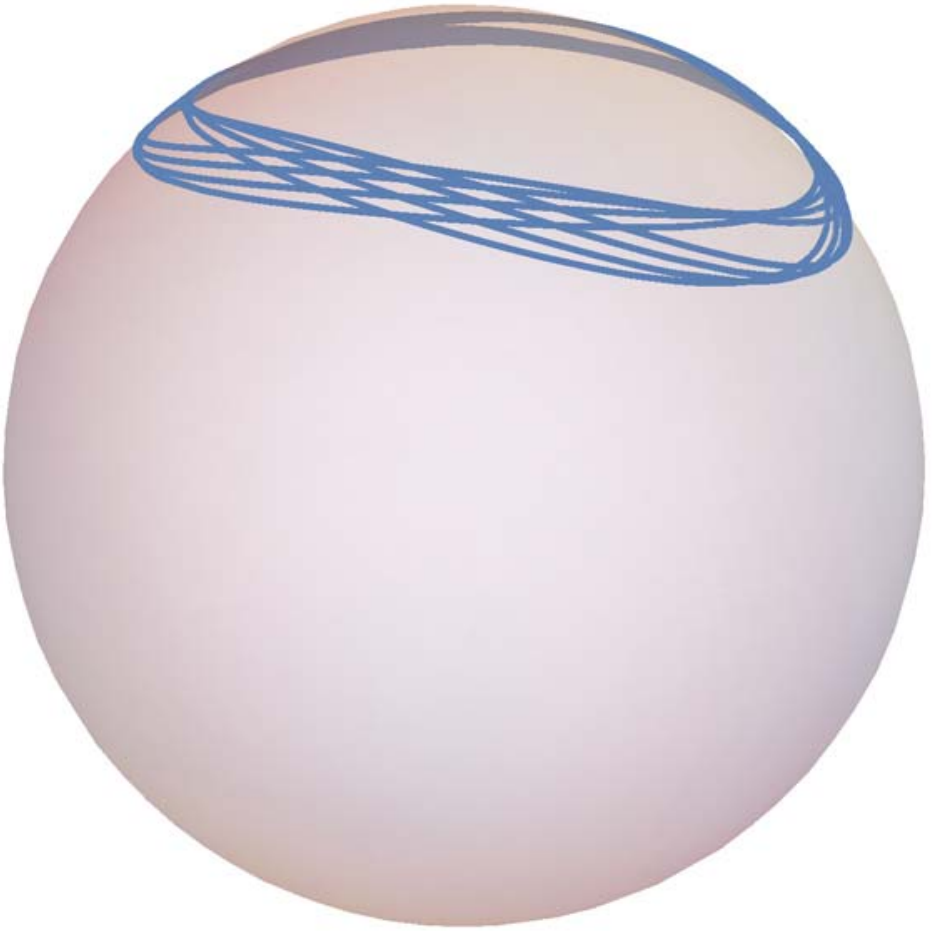} \hspace{0.8cm}   \includegraphics[height=4cm]{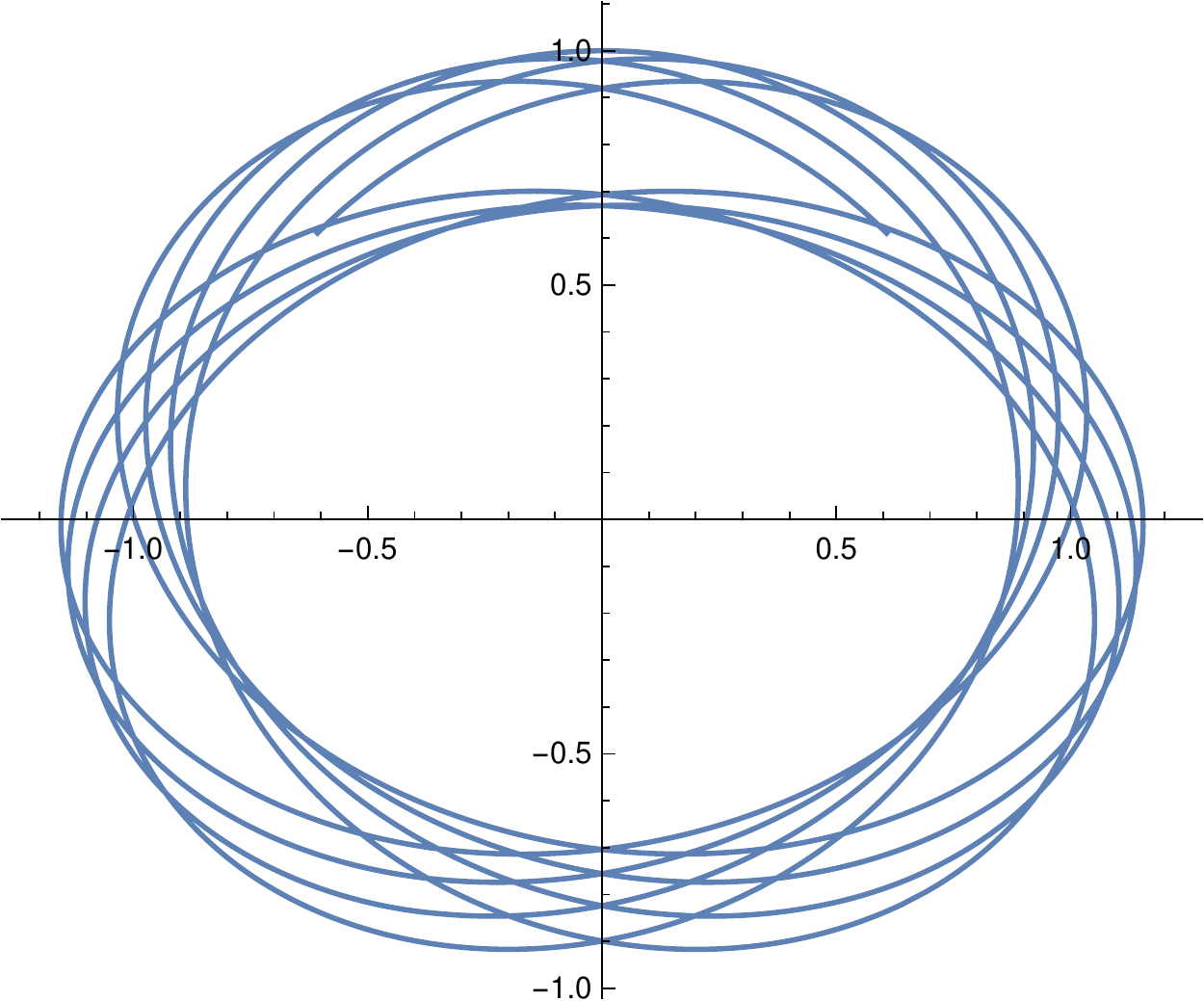}
\caption{An orbit of the Neumann problem and its corresponding planar orbit.}
\end{center}
\label{fig:2}       
\end{figure*}

\subsection{The Killing system}

In the Killing system \cite{Killing1885,Kozlov1992}, see also \cite{Gonzalez} and references therein, one massive particle is forced to move on an $S^2$-sphere of radius $R$ under the action of a gravitational field created by two (e.g. attractive, $\gamma_1>0$, $\gamma_2>0$) Keplerian centers. Fixing the centers in the above defined $F_1$ and $F_2$ points the potential energy reads:
 \[
{\cal U}(\theta_1,\theta_2)=-\frac{\gamma_1}{R}{\rm cotan}\, \theta_1-\frac{\gamma_2}{R}{\rm cotan}\, \theta_2 \quad ,
\]
and thus the test mass feels the presence of two attractive centers in the North hemisphere and two (repulsive) ones located at the antipodal points in the South hemisphere with identical strengths. In sphero-conical coordinates the potential energy is written with two different expressions depending on the hemisphere that it is considered. In both cases ${\cal U}_{\pm}(U,V)$ is of Liouville Type I in $S^2$ form (\ref{typeIS2}), with:
\[
F_{\pm}(U)=\mp  \frac{\gamma_1+\gamma_2}{R}U\sqrt{1-U^2} \quad , \quad G(V)=-\frac{\gamma_1-\gamma_2}{R} V\sqrt{1-V^2} \quad .
\]
Applying the general procedure explained above the dynamics in the $S^2_+$ hemisphere can be described by the planar potential:
\[
{\cal V}_+(x_1,x_2)= -\frac{\gamma_1}{\sigma^2\, \sqrt{(x_1-\frac{R\bar{\sigma}}{\sigma})^2+x_2^2}}-\frac{\gamma_2}{\sigma^2 \sqrt{(x_1+\frac{R\bar{\sigma}}{\sigma})^2+x_2^2}}
\]
that corresponds to the problem of two attractive centers in ${\mathbb R}^2$. In a parallel way, the problem in the South hemisphere is orbitally equivalent to the planar problem:
\[
{\cal V}_-(x_1,x_2)= \frac{\gamma_2}{\sigma^2\, \sqrt{(x_1-\frac{R\bar{\sigma}}{\sigma})^2+x_2^2}}+\frac{\gamma_1}{\sigma^2 \sqrt{(x_1+\frac{R\bar{\sigma}}{\sigma})^2+x_2^2}}\  ,
\]
i.e. the planar potential of two repulsive centers where the roles of the points $(\pm \frac{R\bar{\sigma}}{\sigma}, 0)$, and thus the strengths of the centers in modulus, are
interchanged with respect to the attractive potential ${\cal V}_+(x_1, x_2)$ in $\Pi_+(S^2_+)$.

In \cite{Gonzalez} a complete analysis of the different types of orbits for this problem is performed, including the integration and inversion of the involved elliptic integrals that lead to explicit expressions in terms of Jacobi elliptic functions for all the available regimes in the bifurcation diagram. Two examples of planetary type orbits are represented in Figure 4. In Figure 5 we can see their corresponding orbits in the projected planar systems.

\begin{figure*}
\begin{center}
  \includegraphics[height=5cm]{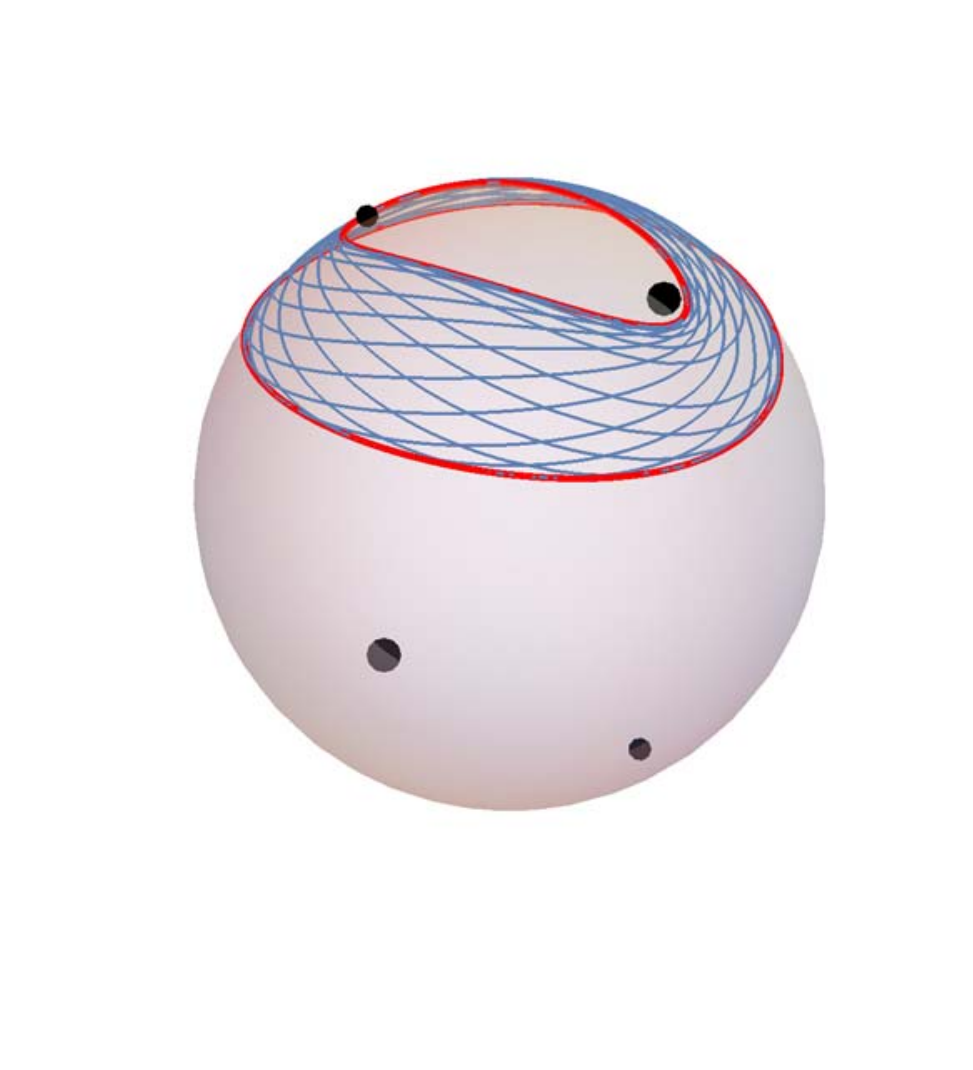}   \includegraphics[height=5cm]{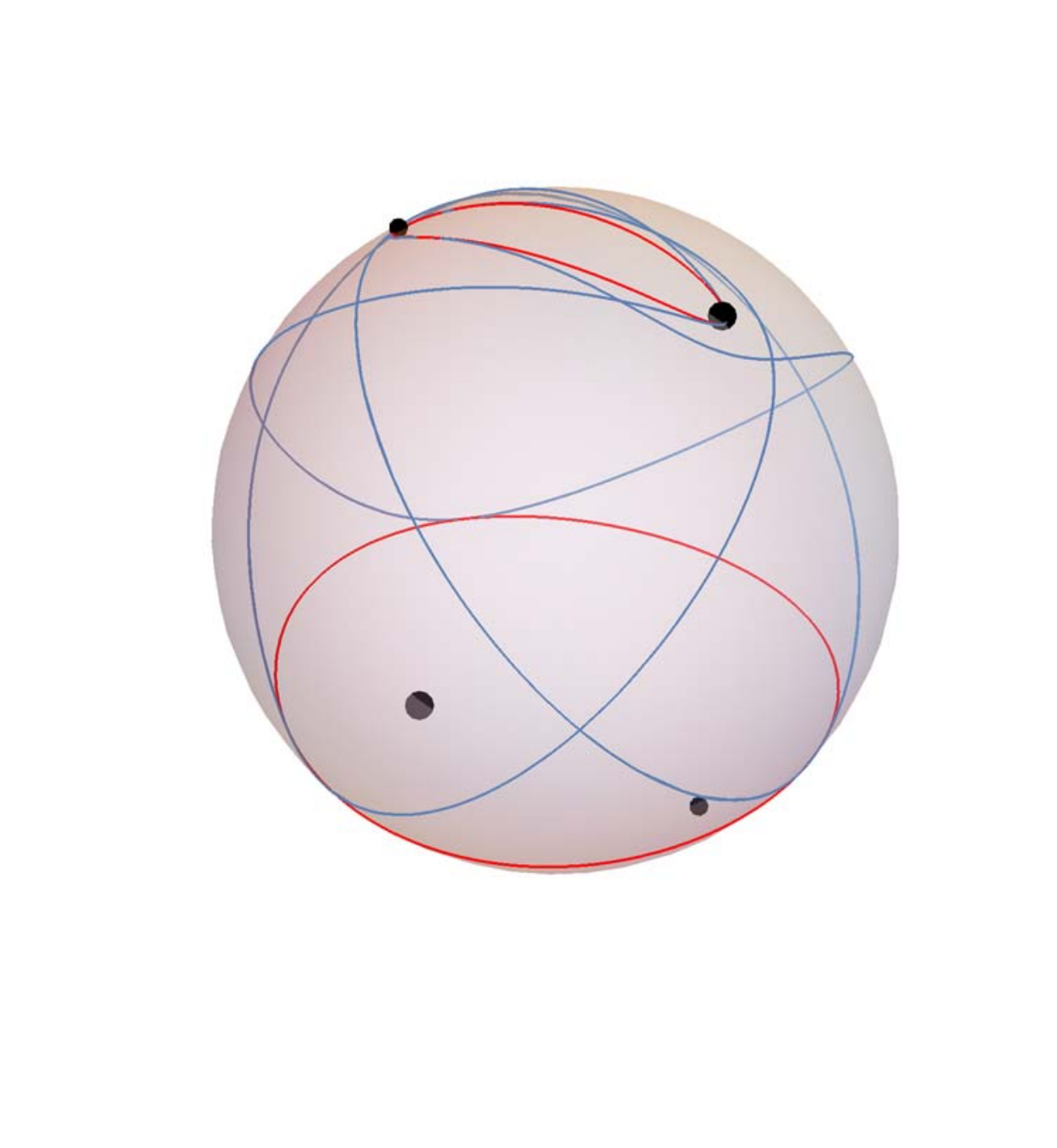}
\caption{a) Planetary orbit in $S^2_+$. b) Closed orbit in $S^2$ that crosses the equator.}
\end{center}
\label{fig:2}       
\end{figure*}

\begin{figure*}
\begin{center}
\includegraphics[height=2.7cm]{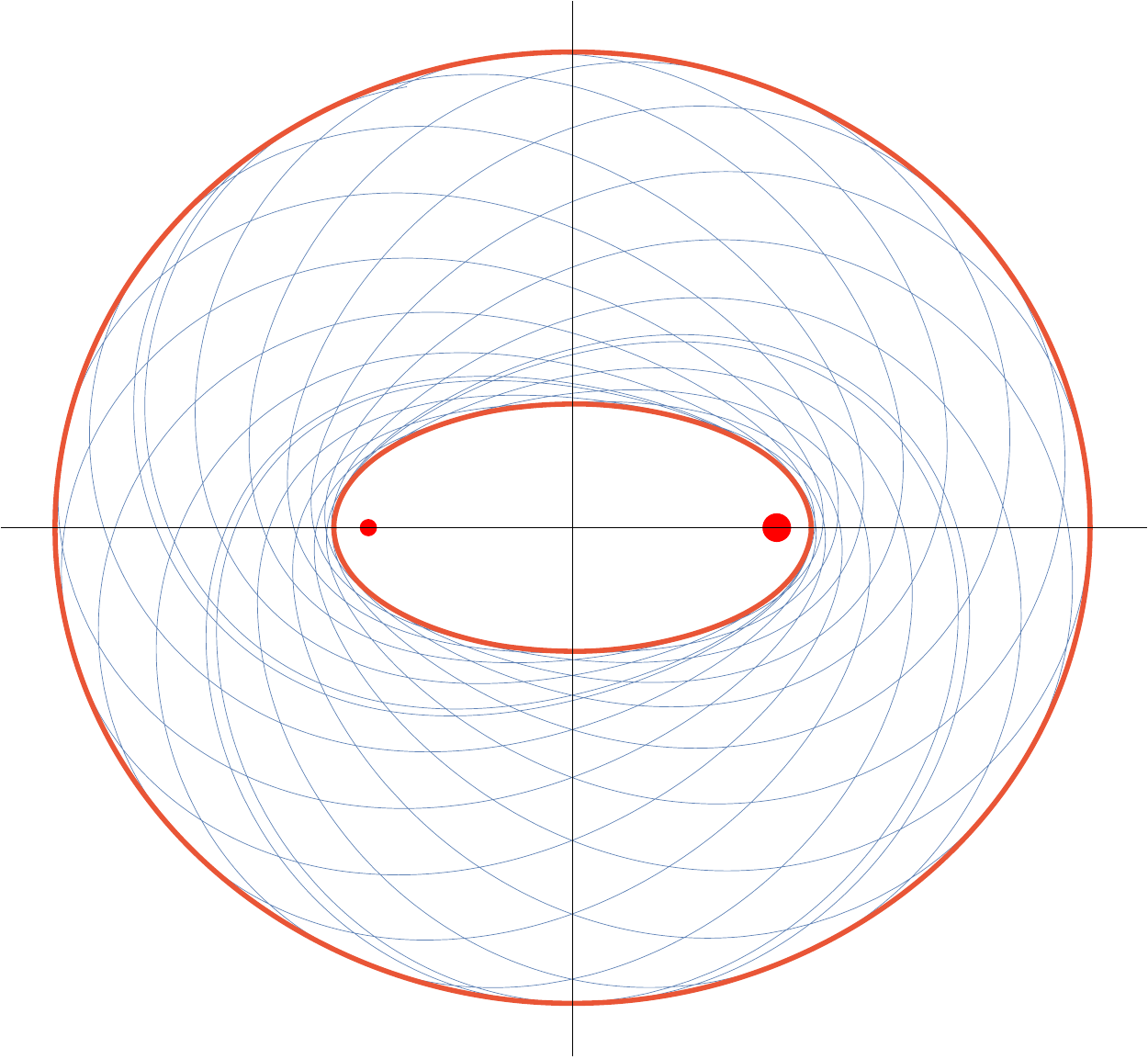} \hspace{0.5cm} \includegraphics[height=2.7cm]{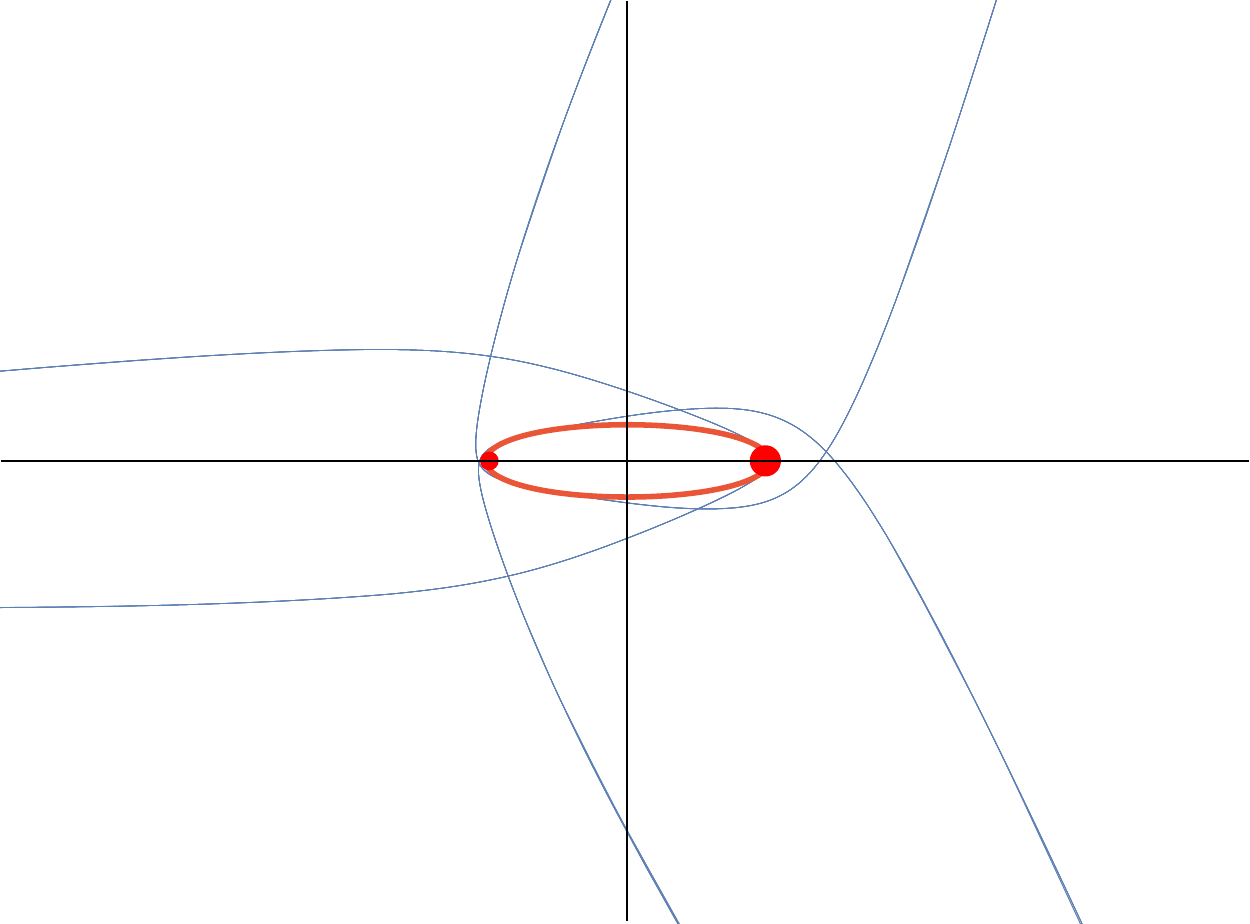}\includegraphics[height=2.7cm]{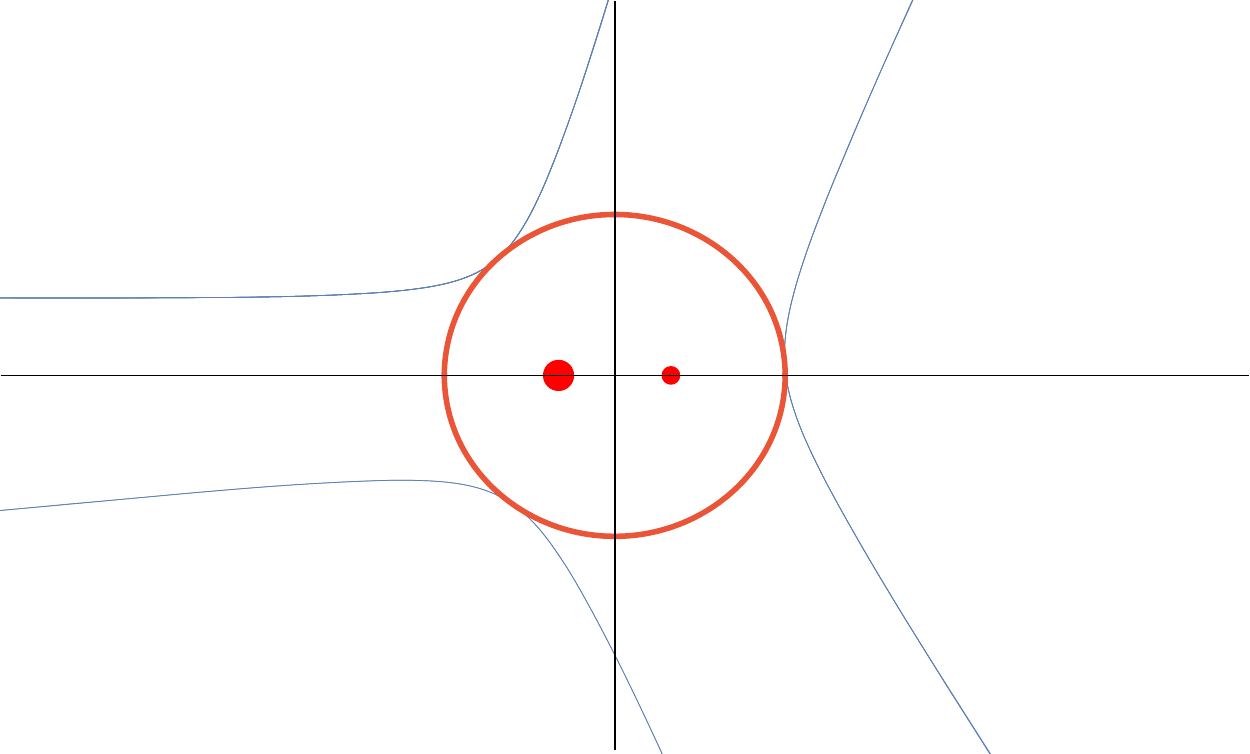}
\caption{a) Planetary orbit projected in ${\mathbb R}^2$. b) and c) Projections of the closed orbit on $\Pi_+(S^2_+)$ and $\Pi_-(S^2_-)$ respectively.}
\end{center}
\label{fig:2}       
\end{figure*}

\subsection{Inverse gnomonic projection of the Garnier system from ${\mathbb R}^2$ to $S^2$}
The Garnier system \cite{Garnier,Perelomov} corresponds to a planar anharmonic oscillator which is isotropic in the quartic power of the distance to the center but anisotropic in the quadratic term. Using non-dimensional coordinates and couplings the potential energy is defined to be:
\[
{\cal V}(x_1,x_2)= \frac{1}{2}\left(x_1^2+x_2^2-1\right)^2+\frac{a^2}{2}x_2^2 \quad , \qquad 0<a<1 \quad .
\]
Changing to Euler elliptic variables (\ref{elliptic}) it is easily seen that it is a Liouville Type I system in ${\mathbb R}^2$ since the potential energy takes the form (\ref{sepellip}) where:
\[
f(u)=\frac{a^4}{2} ( u^2-1) \left( u^2-\frac{1}{a^2}\right)^2\ , \quad g(v)=\frac{a^4}{2} ( 1-v^2) \left( v^2-\frac{1}{a^2}\right)^2 \quad .
\]
The quadratures are thus
\[
\varsigma-\varsigma_0=\mp  \, a \int_1^u \, \frac{d z}{\sqrt{(z^2-1)P_6(z) }} \  , \quad \varsigma-\varsigma_0=\mp  a \int_{-1}^v \, \frac{d z}{\sqrt{(1-z^2)\tilde{P}_6(z)}} \quad ,
\]
where the sixth order polynomials read:
\[
P_6(u)=-f(u)+h u^2-\lambda\quad ,\qquad \tilde{P}_6(v)=-g(v)-hv^2+\lambda \quad .
\]
The change of integration variable: $z^2 = \bar{z}$, renders both integrals to identical canonical form:
\[
\int \frac{d\bar{z}}{\sqrt{2 \bar{z} (1-\bar{z}) (a^4 \bar{z}^3-a^2(a^2+2) \bar{z}^2+(1+2a^2-2h)\bar{z} +2\lambda-1)}}
\]
i.e., they are hyper-elliptic integrals of genus 2.

The inverse gnomonic projection leads us to the Liouville Type I separable system in $S^2_+$ characterized by the rational functions:
\[
F(U)=\frac{(U^2-\bar{\sigma}^2) (1-2U^2)^2}{2(1-U^2)^2}  \  , \quad  G(V)=\frac{(\bar{\sigma}^2-V^2) (1-2V^2)^2}{2(1-V^2)^2}
\]
where, in this non dimensional setting, we have identified the parameters in the form: $a=\frac{\bar{\sigma}}{\sigma}$.

The corresponding potential in terms of $(X,Y,Z)\in S^2_+$ is:

\[
{\cal U}(X,Y,Z)=\frac{1}{2\sigma^2} \left( \frac{1- \bar{\sigma}^2 X^2}{Z^4}- \frac{1+3\sigma^2}{Z^2}\right)
\]
that is singular in the Equator. Thus in this case, even if we extend ${\cal U}(X,Y,Z)$ to the whole $S^2$ sphere, the orbits cannot cross the Equator, and unbounded planar orbits are mapped into spherical trajectories that approach the Equator asymptotically.

\section{Summary and further comments}

In this report we have analyzed separable classical Hamiltonian systems in an unified way. We have focused in systems of two degrees of freedom for which the configuration space is either an $S^2$ sphere or the Euclidean plane ${\mathbb R}^2$.
In the first case, that we denote as Liouville Type I in $S^2$, we have selected those systems for which the Hamilton-Jacobi equation is separable in
sphero-conical coordinates. In the planar case the separability
of the HJ equation is demanded in Euler elliptic coordinates, thus restricting ourselves to Liouville Type I systems
in ${\mathbb R}^2$.

The main contribution in this essay is the construction of a bridge between Liouville Type I systems respectively in
$S^2$ and ${\mathbb R}^2$. The path is traced following the gnomonic projection from both the North and South hemi-spheres to the plane.
The idea is inspired by the connection between the two Keplerian center problem respectively in $S^2$ and ${\mathbb R}^2$ established in \cite{Al1,Al2,Borisov2007}. We provide a geometric structure to the Borisov-Mamaev map which allows to extend the idea to any Liouville Type I system. As particular cases we construct the bridge between the Neumann problem and its partner in
the plane, besides reconstructing the Borisov-Mamaev map betwen the Killing problem, two Keplerian centers in $S^2$,
and the Euler problem, two Keplerian centers in ${\mathbb R}^2$, in this geometric setting. Moreover, we also consider a distinguished Liouville Type I system in ${\mathbb R}^2$, the Garnier system and its mapping back in $S^2$ by using the inverse of the gnomonic projection. A remarkable fact emerges: the two center problems in $S^2$ and ${\mathbb R}^2$ exhibit
separable potentials in terms of either trigonometric or polynomial functions but identical, up to a constant, strengths: in both manifolds Keplerian potentials arise.

The results of this work can be extended to the Quantum framework. It would be very interesting to analyze the relation between separable Schr\"odinger equations in $S^2$ and the corresponding projected equations in ${\mathbb R}^2$. Connecting paths between the classical and quantum worlds are provided by the WKB quantization procedure.

Finally, it is adequate to remind that the search for solitary waves arising in $(1+1)$-dimensional relativistic scalar field theories is tantamount to solve an analogue mechanical system. In this framework, the application of the equivalence between separable systems in $S^2$ and ${\mathbb R}^2$ could be a fruitful source of information about the links between solitary waves in non-linear and linear sigma models, \cite{Alonso,Alonso1,Alonso2,Alonso3}.

\section*{Acknowledgements}
The authors thank the Spanish Ministerio de Econom\'{\i}a y Competitividad (MINECO) for financial support under grant MTM2014-57129-C2-1-P and the Junta de Castilla y Le\'on for the grant VA057U16.

\end{document}